\documentclass[12pt,showpacs,preprintnumbers,superscriptaddress,nofootinbib]{revtex4-1}
\usepackage{amsmath}
\usepackage{amsfonts}
\usepackage{amssymb}
\usepackage{amstext}
\usepackage{graphicx}
\usepackage{mathtools}
\usepackage{float}
\usepackage{subcaption} 

\begin{document}

\title{ \boldmath
	Bounds on Dipole Moments of hidden Dark Matter  through kinetic mixing}

\author{D. T. Binh}
\email{dinhthanhbinh3@duytan.edu.vn}
\affiliation{Institue of Theoretical and Applied Research,
	Duy Tan University, Hanoi 100000, Vietnam}
\affiliation{ Faculty of Natural Science, Duy Tan University, Da Nang 550000, Vietnam}

\author{ V. H. Binh}
\email{vhbinh@iop.vast.ac.vn}
\affiliation{Institute of Physics, Vietnam  Academy of Science and Technology, 10 Dao Tan, Ba Dinh, Hanoi, Vietnam}	
\author{ H. N. Long}
\email{hoangngoclong@tdtu.edu.vn}
\affiliation{Theoretical Particle Physics and Cosmology Research Group, Advanced Institute for Materials Science, Ton Duc Thang University, Ho Chi Minh City, Vietnam}
\affiliation{Faculty of Applied Sciences, Ton Duc Thang University, Ho Chi Minh City, Vietnam}
\date{\today}

\begin{abstract}
	The existence of dark sectors, consisting of  weakly-coupled particles that do not interact with the known Standard Model forces, is theoretically and phenomenologically motivated. The hidden particles are candidates for Dark Matter and  can interact with photon through electric dipole moment (EDM) and  magnetic dipole moment (MDM). We investigate the possibility a hidden sector's Dark Matter which is charged under a hidden $U(1)_X$ gauge symmetry can interact with photon at loop level. We evaluate the  scattering cross section of hidden Dirac fermion with nuclei and set bounds for dipole moment.  Using the results of the XENON1T experiment for direct detection of Dark Matter, we get  bounds of electromagnetic dipole moment $(\mu_\chi)$ for mass $m_\chi=100$ GeV :   $ 1.93448 \times 10^{-8}\mu_B \leq \mu_\chi \leq 1.9496 \times 10^{-8}\mu_B$ and  electric dipole moment $(d_\chi)$: $ 3.3204 \times 10^{-23}e.cm \leq d_\chi \leq 3.3464 \times 10^{-23}e.cm$. Using the condition of the existence of dipole moment we constraint the kinetic mixing parameter   $ 3\times 10^{-3} \leq \epsilon \leq 10^{-2}$ and the mass of the hidden $U(1)_X$ gauge boson  to be in the range of 5 GeV $\leq m_X \leq$ 9 GeV.  Our results complement previous works and are within detection capability of LHC. 
\end{abstract}

\maketitle

\section{Introduction}
\label{intro}

Even though there exits evidence of Dark Matter (DM) \cite{WMAP,Obv-Sig-DM1,Obv-Sig-DM2,Obv-Sig-DM3,Obv-Sig-DM4}, the origin and properties of DM is still mysterious. It is clear that the origin and the properties of DM cannot be addressed in the framework of the Standard Model  (SM) hence the need for the extension of the SM.  There are many possible scenarios for DM candidates in beyond SM. To date, Weakly Interacting Massive Particles( WIMPs)\cite{WIMP-1} and QCD axion \cite{Axion-DM} are the most promising candidates for DM. However, recently  the hidden sector of SM is gaining interests \cite{Abelian-Kinetic-Mixing,DarkSM-1,DarkSM-2,DarkSM-3,DarkSM-4}. In this direction, there exits a hidden sector with different quantum number with SM. This hidden sector includes not only DM particles but also possible hidden force \cite{Hidden-Force-1} that allows DM particles can interact with itself \cite{Self-Interact} and with SM matter through a portal \cite{Hidden-Portal,Higgs-Portal}. 
The  natural way of coupling the SM fields to the dark sector is via the kinetic mixing operator. The kinetic mixing term can be Abelian \cite{Abelian-Kinetic-Mixing}  or nonAbelian \cite{DarkSM-1,DarkSM-2,DarkSM-3,DarkSM-4}. In case of Abelian, the mixing term has the form
\begin{equation}
\mathcal{L} \supset \epsilon F^{\mu \nu}X_{\mu \nu} 
\end{equation}
where $\epsilon$ is a dimensionless parameter. $F^{\mu \nu}$ and $X_{\mu \nu}$ are the field strength of the $U(1)_Y$ gauge field $B_\mu $ and $U(1)_X$ gauge field $X_{\mu }$ respectively. 

The one-loop level result of $\epsilon $  for kinetic mixing with $U(1)_Y$ is \cite{mix-term-1,mix-term-2}: 
\begin{equation}
\epsilon \approx \frac{e g_X}{16 \pi^2}  \log{\frac{m}{\Lambda}}
\end{equation}
where  $g_X$ are the gauge couplings of the $U(1)_X$ group and m is the mass of the heavy particle coupled to the new $U(1)_X$ and $U(1)_Y$. The estimated value of $\epsilon$ is $\epsilon \in (10^{-12},10^{-3})$ \cite{mix-term-3}.

One can introduce a massive Dirac fermion which has charge under $U(1)_X$. This Dirac fermion can be candidate for DM. There are works employed this scenario to implement MeV DM \cite{MeV-DM},  10GeV DM through kinetic mixing \cite{10GeV-DM} or a hidden sector through Higgs mixing at TeV scale \cite{TeV-DM}.

Although DM  has zero electric charge it may couple to photons through loops in the form of electric and magnetic dipole moments. If DM has non-zero  electric or magnetic dipole moment then it can scatter with the nuclei  in direct detection experiments such as: XENON10 \cite{XENON10}, XENON100 \cite{XENON100}, XENON1T \cite{XENON1T-Result}, CDMS \cite{CDMS}, DAMA \cite{DAMA}, CoGeNT \cite{CoGeNT}.\\

In this paper we consider the possibility that hidden DM possesses an electric or magnetic dipole moment.  The paper is organized as follows: In section II we review the kinetic mixing in the  $SU(2)_L$ model. In section III we consider the case a Dirac fermion elastically scatter off a nuclei through photon, Z boson and X boson exchanges. We evaluate the scattering events rate. In section IV we presented the analysis with main type of  results:
\begin{itemize}
	\item We find bounds on dipole moment of hidden dark matter using the Dark matter search experiment. 
	\item We find the regions in parameter space with positive signal from the XENON1T experiment. 
\end{itemize}

In section V we summarize our results and our work.

\section{ Kinetic mixing in the $SU(2)_L$ model}
\label{sec2}
In this section we will briefly review the kinetic mixing in the $SU(2)_L$ model proposed as in \cite{Abelian-Kinetic-Mixing}. 
We investigate a hidden sector containing a gauge symmetry $U(1)_X$ and a Dirac fermion
dark matter candidate $\chi$ couple to the Standard Model sector
through kinetic mixing.

The Lagrangian for this model is as followings:

\begin{eqnarray}
{\cal L}&=& {\cal L}_{SM} - \frac{\sin{\epsilon}}{2}  \hat{B}_{\mu\nu} \hat{X}^{\mu\nu} -\frac{1}{4}\hat{X}^{\mu\nu}\hat{X}_{\mu\nu} \nonumber \\
 &-& g_X \hat{X}^\mu \bar{\chi}\gamma_\mu \chi + {1\over2} m_{\hat{X}}^2 \hat{X}^2
  + m_\chi \bar{\chi}\chi,
\end{eqnarray}

where $\mathcal{L}_{SM}$ is the Lagrangian for the Standard Model, $\hat{B}_{\mu \nu}$,  $\hat{X}_{\mu \nu}$ are field strength tensor of $U(1)_Y$ gauge field $B_{\mu}$ and  $U(1)_X$  gauge field $\hat{X}_\mu$ respectively. Note that in this model besides Standard Model parameters, there are four new parameters which are: mixing parameter $\epsilon$, the mass of the new hidden gauge boson $m_{ \hat{X}}$, new coupling constant $g_X$ and the mass of the new hidden  Dirac fermion $m_{ \chi}$. These parameters can be constrained by several experiments.

The kinetic mixing term and mass mixing  terms can be  diagonalized by the  transformations: \cite{Abelian-Kinetic-Mixing}.

\begin{eqnarray} \label{transformation}
 \hat{B} &=& c_{\hat{W}} A - (t_\epsilon s_\xi+ s_{\hat{W}} c_\xi) Z
 + (s_{\hat{W}} s_\xi-t_\epsilon c_\xi) X \,, \nonumber\\
 \hat{W}_3 &=& s_{\hat{W}} A + c_{\hat{W}} c_\xi Z
 - c_{\hat{W}} s_\xi, \nonumber\\
 \hat{X} &=& \frac{s_\xi}{c_\epsilon}Z + \frac{c_\xi}{c_\epsilon} X \,,
\end{eqnarray}

where the angle $\xi$ is determined by:
\begin{equation}
\tan 2\xi = - \frac{ m_{\hat{Z}}^2 s_{\hat{W}} \sin2\epsilon}{m_{\hat{X}}^2 - m_{\hat{Z}}^2 (c^2_\epsilon-s^2_\epsilon s_{\hat{W}}^2) } \,.
\label{tan2theta}
\end{equation}

and $s_{\hat{W}}$ is defined as \cite{Abelian-Kinetic-Mixing}. 

\begin{equation}
\rho=\frac{s_W^2}{s^2_{\hat{W}}}
\end{equation}

The masses of  $X$ and $Z$ gauge bosons can be  redefined as:
\begin{eqnarray}
 m_Z^2 &=& m_{\hat{Z}}^2(1+s_{\hat{W}} t_\xi t_\epsilon)\,  \label{mz} \\
 m_X^2 &=&\frac{m_{\hat{X}}^2}{c_\epsilon^2 (1+s_{\hat{W}} t_\xi t_\epsilon)} \,.
\label{eq:mx}
\end{eqnarray}

in which $t_\xi $ can be evaluated as a function of $r_X\equiv m_X^2/m_Z^2$ as followings:
\begin{eqnarray}
t_\xi &=& - \frac{1}{s_{\hat{W}} t_\epsilon},\label{t_theta1}\\
t_\xi&=& \frac{1-r_X \pm \sqrt{(1-r_X)^2-4s_{\hat{W}}^2
t_\epsilon^2 r_X}}{2s_{\hat{W}} t_\epsilon r_X}\;. \label{t_theta2}
\end{eqnarray}

The full Lagrangian is given in \cite{Abelian-Kinetic-Mixing}, the relevant  terms for this work are:

\begin{equation}\label{LZ}
 \mathcal{L}_Z =  Z_\mu \left[ g^Z_{ffL}\, \bar{f} \gamma^\mu P_L f
                          + g^Z_{ffR}\, \bar{f} \gamma^\mu P_R f
       + g^Z_{\chi \chi}  \bar{\chi}\gamma^\mu\chi \right] 
\end{equation}

\begin{equation}
 \mathcal{L}_X =X_\mu \left[ g^X_{ffL}\, \bar{f} \gamma^\mu P_L f
                + g^X_{ffR}\, \bar{f} \gamma^\mu P_R f
       + g^X_{\chi \chi}  \bar{\chi}\gamma^\mu\chi \right] 
\end{equation}

From this one can obtain the coupling constants:

\begin{eqnarray}
g^Z_{ffL} &=& -{e\over c_{{W}} s_{{W}} }\, c_\xi \,
         \left\{ T_3 \left[1+ {\omega \over 2}\right]
         - Q \left[s_{{W}}^2 + \omega \left( {2 - t_W^2
         \over 2( 1- t_W^2) }\right) \right] \right\}\,, \\
g^Z_{ffR} &=& {e\over c_{{W}} s_{{W}} }\, c_\xi \,
         Q \left[s_{{W}}^2 + \omega \left( {2 - t_W^2
         \over 2( 1- t_W^2) }\right) \right]\,,    \\
g^Z_{\chi \chi} &=& - g_X {s_\xi\over c_\epsilon}\,, \\
g^X_{ffL} &=& - {e\over c_{{W}} s_{{W}} }\, {c_\xi} \left\{
         T_3 \left[ s_W t_\epsilon - t_\xi + {1\over 2}\,
         \omega \left(t_\xi + { s_W t_W^2 t_\epsilon\over 1 - t_W^2} \right) \right]
         \right. \nonumber \\
  &+& \left.  Q \left[ s_W^2 t_\xi - s_W t_\epsilon
         + {1\over 2}\, t_W^2 \omega \left({t_\xi -s_W t_\epsilon \over 1-t_W^2 } \right) \right]
         \right\}\,, \\
g^X_{ffR} &=& -{e\over c_{{W}} s_{{W}} }\, {c_\xi}\, Q \left[ s_W^2 t_\xi - s_W t_\epsilon
         + {1\over 2}\, t_W^2 \omega \left({t_\xi -s_W t_\epsilon \over 1-t_W^2 } \right)
         \right]\,, \\
g^X_{\chi \chi } &=& - g_X {c_\xi\over c_\epsilon}\,, 
\end{eqnarray}

\section{Dipole Moment Interaction of Dark Matter}

\begin{figure}
	\centering
		\includegraphics[scale=.75]{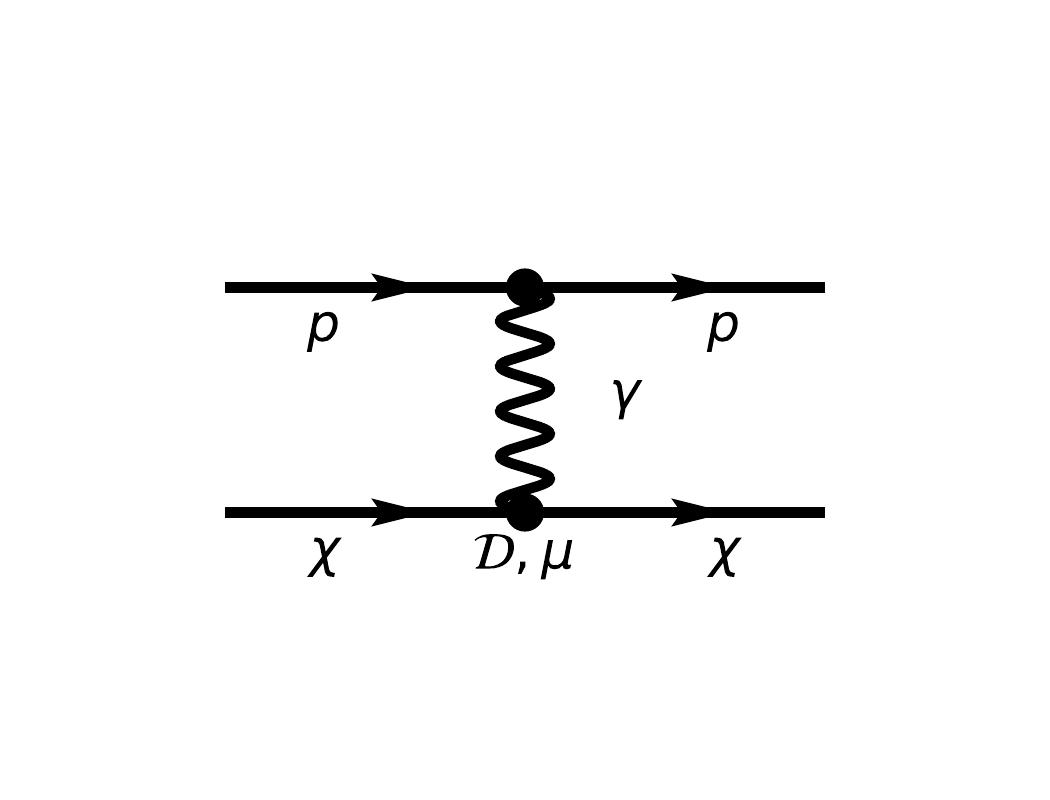}
		\includegraphics[scale=.75]{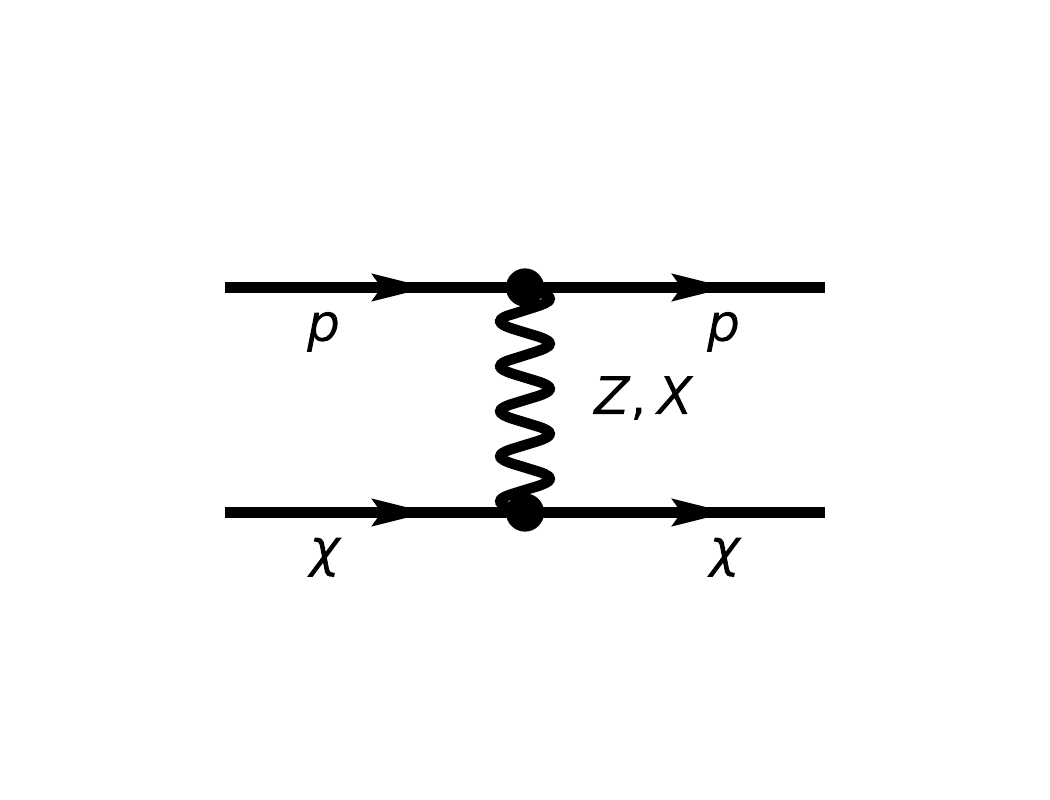}
	\caption{Feynman diagrams of dark matter scattering off a nuclei}
	\label{Nucleon-DM-HS-Scattering}
\end{figure}

The elastic scattering of the DM $\chi$ by a nuclei at  t-channel through  $X$, $Z$ gauge boson exchange and electromagnetic scattering  as in  Figure.\ref{Nucleon-DM-HS-Scattering}. 
The  effective Lagrangian for coupling of a Dirac fermion $\chi$ having an electric dipole moment $d_\chi$ and a magnetic dipole moment $\mu_\chi$ to a electromagnetic field  $F_{\mu \nu}$ is:

\begin{equation}
\mathcal{L}=-\frac{i}{2} \bar{\chi}\sigma_{\mu \nu}(\mu_\chi +\gamma_5 d_\chi)\chi F^{\mu \nu}
\end{equation}



where $\sigma_{\mu \nu}=\frac{i}{2}[\gamma_\mu,\gamma_\nu]$\\

In the low energy limit, the propagator of of Z,X bosons are proportional to  $\approx$ $\frac{-1}{m^2_{Z,X}}$. Hence the scattering through  t-channel can be described by the effective Lagrangian:

\begin{eqnarray}
\mathcal{L}_{eff}^{Z, X}&=&\mathcal{L}^Z_{ff} \mathcal{L}^Z{\bar{\chi}\chi}+\mathcal{L}^X_{ff} \mathcal{L}^X{\bar{\chi}\chi} =\mathcal{C}_f \bar{\psi}\gamma_\mu \psi \bar{\chi}\gamma^\mu \chi
\end{eqnarray}

where

\begin{eqnarray}
 \mathcal{C}_f = {g^Z_\psi (g^Z_{fL}+g^Z_{fR}) \over 2 m_Z^2}
     + {g^X_\psi (g^X_{fL}+g^X_{fR}) \over 2 m_X^2} \,.
\label{eq:bf}
\end{eqnarray}

The coefficient of scattering of dark matter of proton and neutron are:

\begin{eqnarray}
\mathcal{C}_p&=&2\mathcal{C}_u + \mathcal{C}_d \\
\mathcal{C}_n &=& \mathcal{C}_u + 2\mathcal{C}_d 
\end{eqnarray}
 These coefficient can be calculated as given in \cite{Abelian-Kinetic-Mixing}:

\begin{eqnarray}
 \mathcal{C}_p &=& \frac{\hat{g} g_X}{4 c_{\hat{W}} } \frac{c_\xi^2}{c_\epsilon} \frac{t_\xi}{m_Z^2}
         \left[ (1-4s_{\hat{W}}^2)\left(1 - \frac{1}{r_X} \right)
         -3 s_{\hat{W}} \frac{t_\epsilon}{t_\xi} \left(t_\xi^2 +  \frac{1}{r_X}\right) \right]  \nonumber\\
     &\simeq & \frac{e g_X}{4 c_W s_W } \frac{c_\xi^2}{c_\epsilon} \frac{t_\xi}{ m_Z^2} \left\{ (1-4 s_W^2) \left( 1 - \frac{1}{r_X} \right)
 - \frac{ 3}{r_X } \frac{ s_W t_\epsilon}{t_\xi } \right.  \nonumber\\
&-&  \left.    \omega \left[ 3 + \left( 1 - \frac{1}{ r_X} \right)
         \left( \frac{1}{2} + 2 s_W^2 {1+t_W^2\over 1-t_W^2} \right)
         - \frac{1}{r_X} \frac{s_W t_\epsilon}{t_\xi} \frac{3 t_W^2}{2 - 2 t_W^2}
          \right]
          \right\},
         \\
 \mathcal{C}_n &=& -\frac{\hat{g} g_X}{4 c_{\hat{W}} } \frac{c_\xi^2}{c_\epsilon} \frac{t_\xi}{m_Z^2}
         \left[ \left(1 - \frac{1}{r_X}\right)
         + s_{\hat{W}} \frac{t_\epsilon}{t_\xi} \left(t_\xi^2 + \frac{1}{r_X}\right)
         \right] \nonumber\\
     &\simeq & -\frac{e g_X}{4 c_W s_W } \frac{c_\xi^2}{c_\epsilon} \frac{t_\xi}{m_Z^2}
         \left\{ \left[ 1 - \frac{1}{r_X} \left(1 - s_W \frac{t_\epsilon}{t_\xi}\right) \right]
         + \frac{\omega}{2} \left[ 1 + \frac{1}{r_X}
         \left( 1 + \frac{s_W t_W^2 t_\epsilon}{(1-t_W^2) t_\xi } \right) \right] \right\}
         \,.~~~~~~~
\end{eqnarray}

The scattering amplitude is:

\begin{eqnarray}
\mathcal{M}_{fi}&=& \sum_q \bar{u}(k_f)\Gamma_\mu u(k_i) \frac{i g^{\mu \nu}}{q^2} \bar{u}(p_f) (ie_q) \gamma_\nu u(p_i)  \nonumber \\
&+& \sum_{q} \mathcal{C}_i \bar{u}(k_f)\gamma_\alpha u(k_i) \bar{u}(p_f)\gamma^\alpha u(p_i)
\end{eqnarray}

where $q=u,d$ and $\Gamma^\mu$ is:

\begin{equation}
\Gamma^\mu=\frac{i}{2}\left( \mu_\chi \sigma^{\mu \nu} q_\nu + d_\chi \gamma^5 \sigma^{\mu \nu} q_\nu \right)
\end{equation}

Taking summation  over all quark flavor we have

\begin{eqnarray}
\mathcal{M}_{fi}&=& Z(ie) \bar{u}(k_f)\Gamma_\mu u(k_i) \frac{i g^{\mu \nu}}{q^2} \bar{u}(p_f) \gamma_\nu u(p_i)  \nonumber \\
&+& [Z C_p + (A-Z)C_n] \bar{u}(k_f)\gamma_\alpha u(k_i) \bar{u}(p_f)\gamma^\alpha u(p_i)
\end{eqnarray}

\subsection{Constraints from direct searches}

We consider the elastic scattering process of in coming dark matter $\chi$ with velocity $v$ scatter off a nucleus $\mathcal{N}(A,Z)$.

\begin{equation}
\chi + \mathcal{N}(A,Z)_{at rest} \rightarrow \chi + \mathcal{N}(A,Z)_{recoil}
\end{equation}

Let $k_i, p_i$  be the momentum of the incoming dark matter and nucleus. $k_f, p_f$ be the momentum of the outgoing dark matter and nucleus. We have  $k_i=(E^i_\chi, \vec{k}_i)$, $p_i=(E^i_N, \vec{p}_i)$, $k_f=(E^f_\chi, \vec{k}_f)$, $p_f=(E^f_N, \vec{p}_f)$. q is the momentum transfer and  $q=k_i-k_f$.

In center of mass frame, for an elastic collision:  $\vec{k}_i=-\vec{p}_i=\vec{k}_f=-\vec{p}_f=M_{\chi N} \vec{v}$

and 
\begin{equation}
q^2=k_i^2 + k_f^2 -2k_i k_f=   2(M_{\chi N} v)^2(1-\cos \theta)
\end{equation}

where $M_{\chi N}$ is the reduced mass of dark matter and nucleus $M_{\chi N}=\frac{m_\chi m_N}{m_\chi + m_N}$ and $\theta$ is scattering angle between dark matter and nucleus.

The recoil energy which is typical  value of $\sim 1-100$ keV is given by:

\begin{equation}
E_R=\frac{q^2}{2 m_N}
\label{RecoiE}
\end{equation}

The minimal value of velocity of the Dark Matter is:

\begin{equation}
v_{min}=\sqrt{\frac{E_R m_N}{2M_{\chi N}^2}}
\end{equation}

 The average amplitude can be calculated as:

\begin{eqnarray}
 \overline{|\mathcal{M}|^2} &=&  8 m_N^2  m_{\chi }^2 \left[(A-Z) C_n+Z C_p\right]^2  \nonumber \\
&+&  \frac{ Z^2 e^2 m_N \ \mu _{\chi }^2 \left[ 4 m_{\chi }^2  -E_R m_N\right]}{E_R} + \frac{ Z^2 e^2 m_N  d_{\chi }^2 \left[2 m_{\chi }^2 - E_R m_N\right]}{E_R} \nonumber \\
\end{eqnarray}

The differential cross section is:

\begin{equation}
\frac{d \sigma}{d \Omega} \approx \frac{1}{64\pi^2(m_\chi+m_N)^2}\overline{|\mathcal{M}|^2}
\end{equation}

From (\ref{RecoiE}) we have 
\begin{eqnarray}
\frac{d \Omega}{dE_R} &=& 2 \pi \frac{m_N}{|\overrightarrow{k_i}||\overrightarrow{k_f}|}  \nonumber \\
\end{eqnarray}
using  $|\overrightarrow{k_i}|=|\overrightarrow{k_f}|=M_{\chi N}v$ we have: 
\begin{equation}
\frac{d \Omega}{dE_R}=\frac{2\pi m_N}{M^2_{\chi N}v^2}
\end{equation}

\begin{eqnarray}
\frac{d \sigma}{dE_R}&=& \left(\frac{d\sigma}{d\Omega}\right) \left(\frac{d\Omega}{dE_R}\right)=\frac{1}{32 \pi m_\chi^2 m_N v^2} \overline{|\mathcal{M}|^2}
\end{eqnarray}

For the low momentum transfer, the nuclei is not a point like particle. It is necessary to consider the charge distribution of the nuclei. We use the Helm form factor \cite{Helm-form-factor-1,Helm-form-factor-2} as a modification of the electromagnetic form factor of the charge distribution in a nuclei. In this form we have

\begin{eqnarray}
\frac{d \sigma}{dE_R}&=&\frac{F^2(E_R)}{32 \pi m_\chi^2 m_N v^2} \overline{|\mathcal{M}|^2}
\end{eqnarray}
where

\begin{equation}
F(E_R)= \frac{ 3j_1(qr)}{qr}   e^{-q^2s^2}
\end{equation}

with  $q=\sqrt{2m_NE_R}$ is the momentum transferred, $m_N$ is the nucleus mass and 

\begin{eqnarray}
j_1(x)=\frac{\sin x}{x^2}  - \frac{\cos x}{x}
\end{eqnarray}

is the Bessel spherical function of the first kind with $r^2=(1.23A^{1/3}-0.6)^2+\frac{7}{3}(0.52\pi)^2-5s^2$ is an effective nuclear radius and  nuclear skin thickness $s\simeq 0.9$ fm \cite{Helm-form-factor-2} \\

\begin{eqnarray}
\frac{d\sigma}{d E_R}&=&\frac{F^2(E_R)}{32 \pi m^2_\chi m_N v^2} \times (  8 m_N^2  m_{\chi }^2 \left[(A-Z) C_n+Z C_p\right]^2    \nonumber \\
&+&  \frac{ Z^2 e^2 m_N \ \mu _{\chi }^2 \left[ 4 m_{\chi }^2  -E_R m_N\right]}{E_R} + \frac{ Z^2 e^2 m_N d_{\chi }^2 \left[2 m_{\chi }^2 - E_R m_N\right]}{E_R} )\nonumber \\
&=&\frac{F^2(E_R)m_N}{4 \pi  v^2} \left[(A-Z) C_n+Z C_p\right]^2   \nonumber \\
&+& \frac{\alpha Z^2 F^2(E_R) }{8 m^2_\chi v^2} \left( \frac{ \mu _{\chi }^2 \left[ 4 m_{\chi }^2  -E_R m_N\right]}{E_R}  +   \frac{ d_{\chi }^2 \left[ 2 m_{\chi }^2  -E_R m_N\right]}{E_R} \right) 
\end{eqnarray}

where $\alpha= \frac{e^2}{4 \pi} $ is fine structure constant.

The event rate $\frac{dR}{dE_R}$ ($\frac{events}{kg.day.keV}$) is :

\begin{eqnarray}
\frac{d R}{dE_R} &=& \frac{\rho_0}{m_\chi m_N}\int  \left[ 
 \frac{F^2(E_R)m_N}{4 \pi  v^2} \left[(A-Z) C_n+Z C_p\right]^2  \right. \nonumber \\
  &+&  \left. \frac{\alpha Z^2 F^2(E_R) }{8 m^2_\chi v^2} \left( \frac{ \mu _{\chi }^2 \left[ 4 m_{\chi }^2  -E_R m_N\right]}{E_R}  +   \frac{d_{\chi }^2 \left[ 2 m_{\chi }^2  -E_R m_N\right]}{E_R} \right) \right]  v f(v) dv
\end{eqnarray}

where $\rho_0$ is local dark matter density  ($\rho_0=0.46^{+0.07}_{-0.09}$ $Gev/c^2/ cm^{3} $) \cite{DM-Density} and  $f(v)$  is  the dark matter velocity distribution in the detector reference frame.

The differential event rate is then calculated  as:

\begin{eqnarray}
\frac{d R}{dE_R} &=& \frac{\rho_0}{m_\chi m_N} \left[ 
\frac{F^2(E_R)m_N}{4 \pi  } \left[(A-Z) C_n+Z C_p\right]^2  \right. \nonumber \\
&+&  \left. \frac{\alpha Z^2 F^2(E_R) }{8 m^2_\chi } \left( \frac{ \mu _{\chi }^2 \left[ 4 m_{\chi }^2  -E_R m_N\right]}{E_R}  +   \frac{ d_{\chi }^2 \left[ 2 m_{\chi }^2  -E_R m_N\right]}{E_R} \right) \right] \int \frac{ f(v)}{v}  dv
\end{eqnarray}

The velocity distribution function $f(v)$ has the Maxwellian distribution  \cite{V-distribution}.

\begin{equation}
f(v)=\frac{v}{v_E v_0 \sqrt{\pi}} \left[   e^{ -\frac{(v-v_E)^2}{v_0^2} } - e^{ -\frac{(v+v_E)^2}{v_0^2}}  \right]
\end{equation}

where   $v_0=220 km/s$  is the  circular velocity of the Sun around the Galatic center and $v_E \sim v_0$ is the velocity of the Earth to the Sun. The average velocity of $v_E=232 km/s$ 
Taking integral the event rate per unit mass and unit time is:

\begin{eqnarray}
R &=& \frac{\rho_0}{m_\chi m_N} \int_{E_{R,min}  }^{ E_{R,max}} dE_R \left[ 
\frac{F^2(E_R)m_N}{4 \pi  } \left[(A-Z) C_n+Z C_p\right]^2  \right. \nonumber \\
&+&  \left. \frac{\alpha Z^2 F^2(E_R) }{8 m^2_\chi } \left( \frac{ \mu _{\chi }^2 \left[ 4 m_{\chi }^2  -E_R m_N\right]}{E_R}  +   \frac{ d_{\chi }^2 \left[ 2 m_{\chi }^2  -E_R m_N\right]}{E_R} \right) \right] \nonumber \\
&\times & \frac{1}{2v_E} \left[ erf\left( \frac{v_{min}+v_E}{v_0}  \right) -erf\left( \frac{v_{min}-v_E}{v_0}  \right)    \right]
\label{EventRate}
\end{eqnarray}

where the error function is defined as:

\begin{equation}
erf(a)= \frac{2}{\sqrt{\pi}}\int_0^a  e^{-x^2} dx
\end{equation}

and  $v_{min}=\sqrt{\frac{E_R m_N}{2 M_{\chi N} ^2}}$ with $M_{\chi N}=\frac{m_N m_\chi }{m_N+m_\chi }$.

\section{Numerical Analysis}

In finding the dipole moment of the Dirac dark matter, we evaluate the event rate as the function of new parameters $\sin{\epsilon},  g_X, m_X, m_\chi $.  In \cite{Abelian-Kinetic-Mixing}, using result in  muon $g-2 $, atomic parity variation, $\rho$ parameter and Electro-Weak Precision Test (EWPT) authors can put some constraints on these four parameters specially the parameters are constrained  as : $ 0.003 \leq \sin \epsilon \leq 0.1 $ if $m_X<m_Z $ and  $ 0.05 \leq \sin \epsilon \leq 0.4 $ if $m_X>m_Z $. Combine with previous constraints \cite{mix-term-1,mix-term-2,mix-term-3,mix-term-4}, we focus on region  where  $3 \times 10^{-3} \leq \epsilon \leq 10^{-2}$ and $m_X \leq m_Z$.

The recoil energy in direct detection experiment typically from $1 \sim 100$keV with event rate $\frac{dR}{dE_R} \in [10^{-7}, 10^{-2}] $. In the kinetic mixing of the $SU(2)_L$ model, besides photon, dark matter can scatter off nuclei through Z boson and new hidden gauge boson. The scattering through photon is proportional to the dipole moment of the dark matter.  Therefore using this condition we can set lower bounds for the dipole moment of the hidden DM  particle. As mentioned above, the hidden Dirac fermion can interact with photon at loops level. Therefore there exits dipole moments for this hidden fermion and the value is greater than zero.

\begin{figure}[ht]
	\centering
	\begin{subfigure}[b]{0.45\textwidth}
		\includegraphics[width=\textwidth]{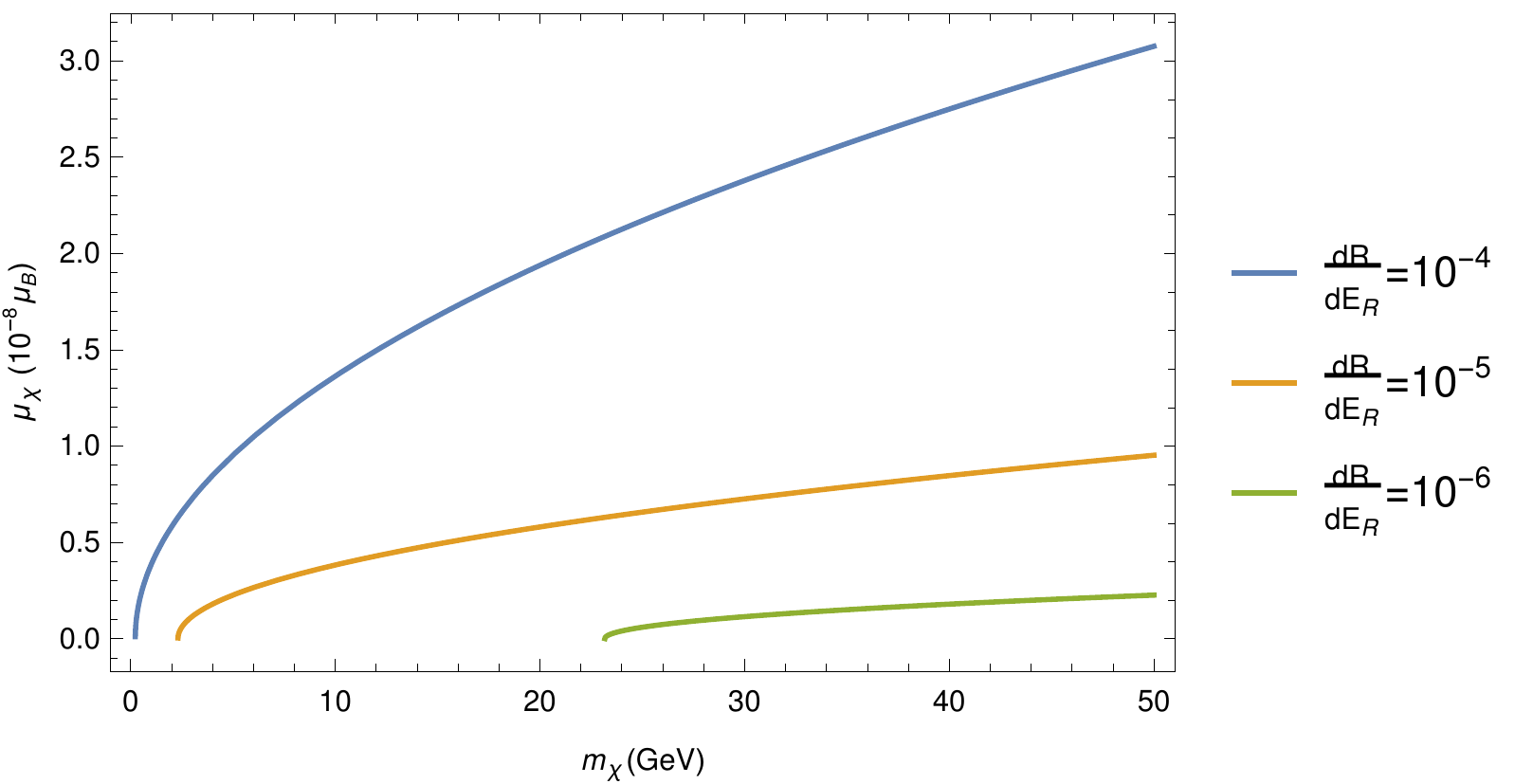}
		\caption{}
		\label{Fig2a}
	\end{subfigure}
	\begin{subfigure}[b]{0.45\textwidth}
		\includegraphics[width=\textwidth]{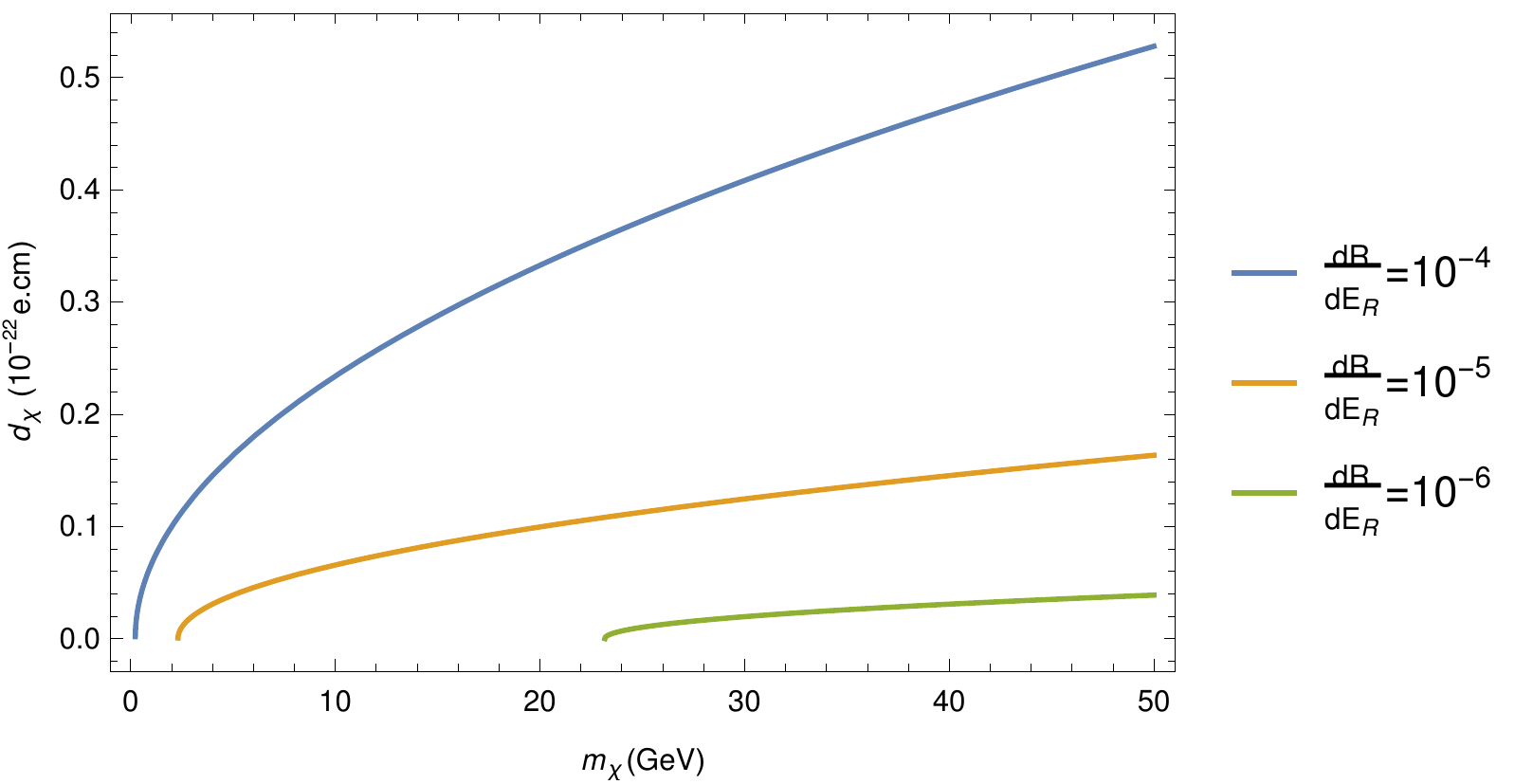}
		\caption{}
		\label{Fig2b}
	\end{subfigure}
\caption{Magnetic dipole moment and electric dipole moment as function of Dirac dark matter $m_{\chi}$ for recoil energy $E_R=30$keV and cpd/keV= $10^{-4},  10^{-5}, 10^{-6}$ }
\label{Fig2}
\end{figure}

We will first evaluate the magnetic moment of $\chi$ hidden Dirac fermion by setting $d_{\chi}=0$ in the expression for $\frac{dR}{dE_R}$ Figure 2(a). The choice of of recoil energy $E_R=30$ keV for Xenon nuclei $\prescript{131}{54}{\mathbf{Xe}}
$. The mass of X boson $m_X=10 $GeV is chosen with mixing parameters $\epsilon=0.003$.  We evaluate the value of $\mu_{\chi }$ as function of the $m_{\chi}$ in three case  $\frac{dR}{dE_R}=10^{-4}, 10^{-5}, 10^{-6}$ respectively. Using the condition that the magnetic dipole moment must be positive we can set lower bound for the mass of the hidden Dirac particle.  In the case $\frac{dR}{dE_R}=10^{-5}$ the lower bound of the mass is $2.32$ GeV \ref{Fig2}(a). If $\frac{dR}{dE_R}=10^{-6}$ than this value is $23.17$ GeV.

\begin{figure}[ht]
	\centering
	\begin{subfigure}[b]{0.45\textwidth}
		\includegraphics[width=\textwidth]{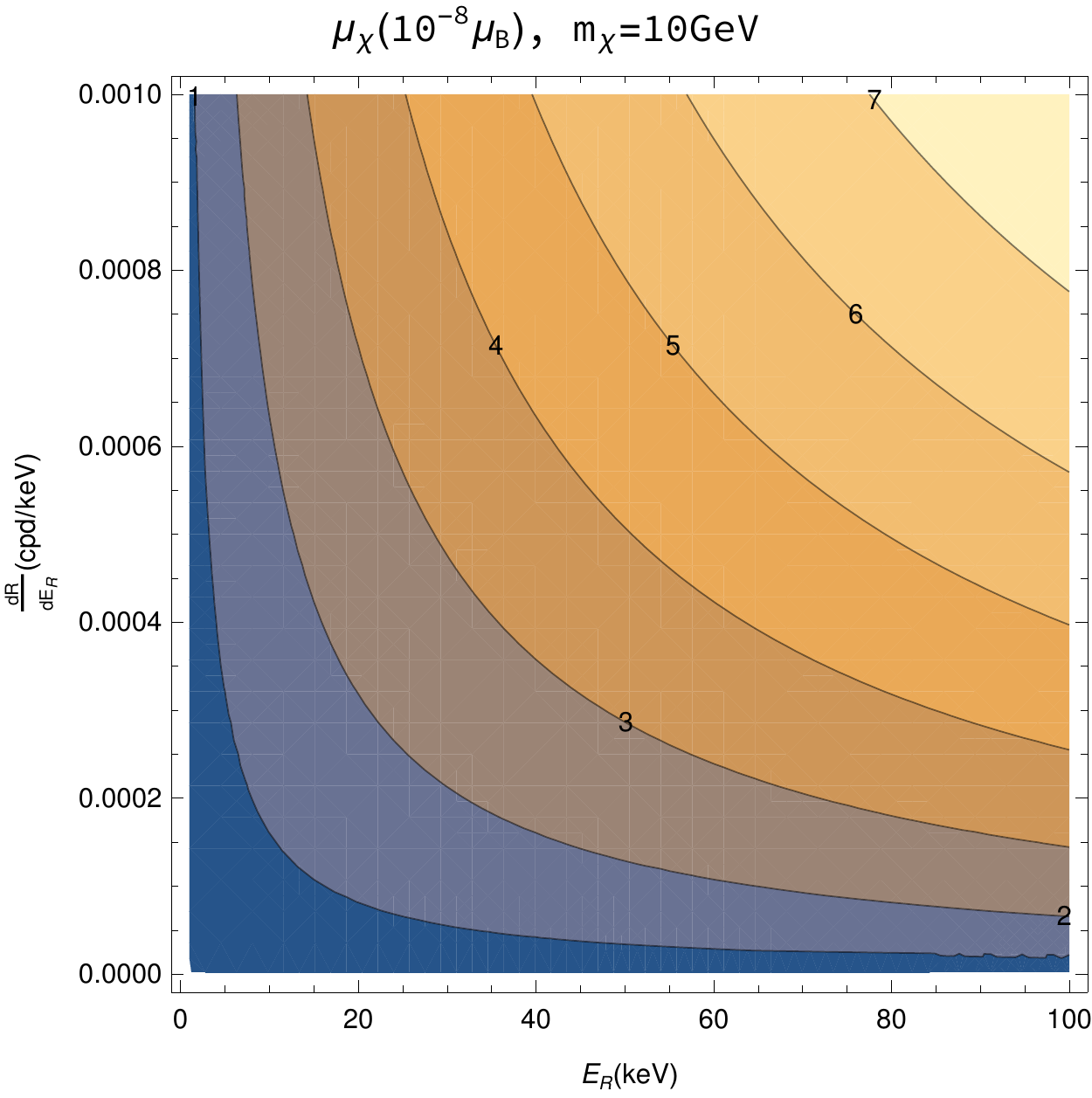}
		\caption{}
		\label{Fig3a}
	\end{subfigure}
	\begin{subfigure}[b]{0.45\textwidth}
		\includegraphics[width=\textwidth]{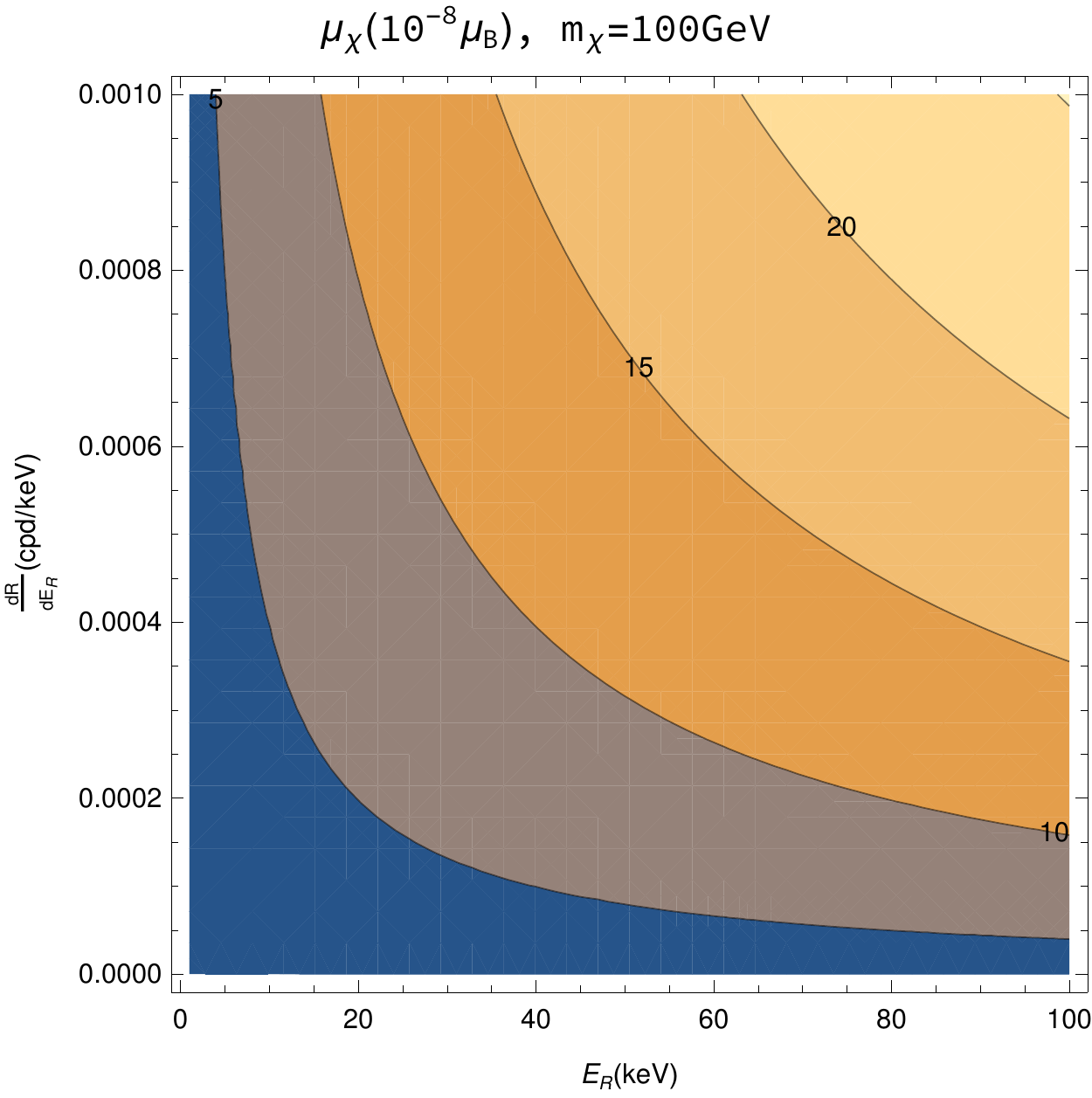}
		\caption{}
		\label{Fig3b}
	\end{subfigure}
	\caption{Magnetic dipole moment as as function of events rate/keV and the recoil energy for two values of $m_{\chi}=10$GeV (a) and $m_{\chi}=100$GeV (b)   }
	\label{Fig3}
\end{figure}

In Figure. \ref{Fig3} we investigate the magnetic dipole moment as function of events rate/keV and the recoil energy for two values of $m_{\chi}=10,100$GeV. If the mass of the hidden Dirac fermion at order $\mathcal{O}(10 )$GeV the magnetic dipole moment $\mu_\chi $ has lower bounds of $1 \times10^{-8} \mu_B$ and $5 \times10^{-8} \mu_B$ if the mass $m_{\chi}=100$ GeV.  We observe that the variation of magnetic dipole moment with respect to the event rate  is decreasing with the increase of the recoil energy $E_R$ and approximate constant with the value of $E_R\geq 40keV$

\begin{figure}[ht]
	\centering
	\begin{subfigure}[b]{0.45\textwidth}
		\includegraphics[width=\textwidth]{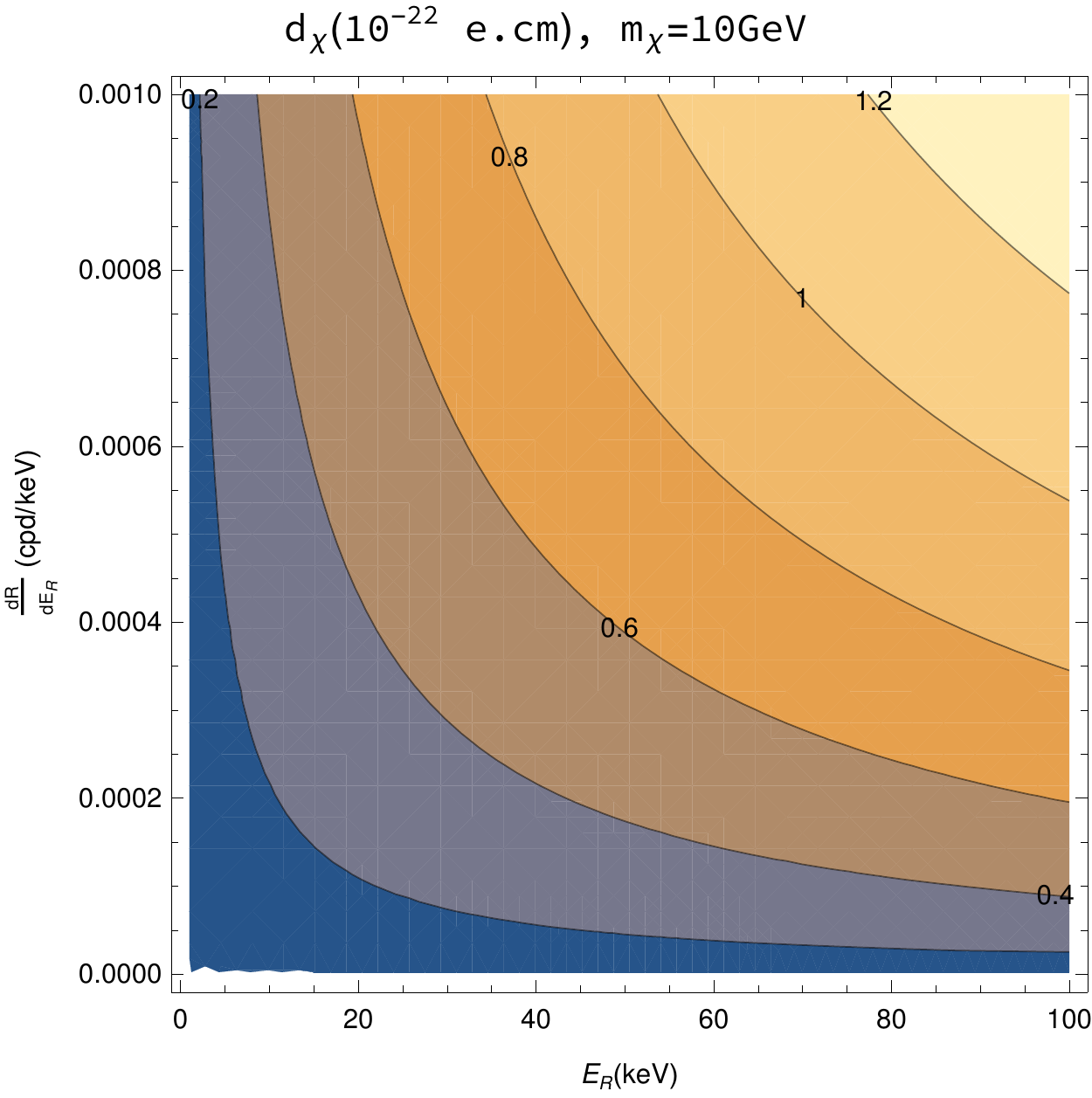}
		\caption{}
		\label{Fig4a}
	\end{subfigure}
	\begin{subfigure}[b]{0.45\textwidth}
		\includegraphics[width=\textwidth]{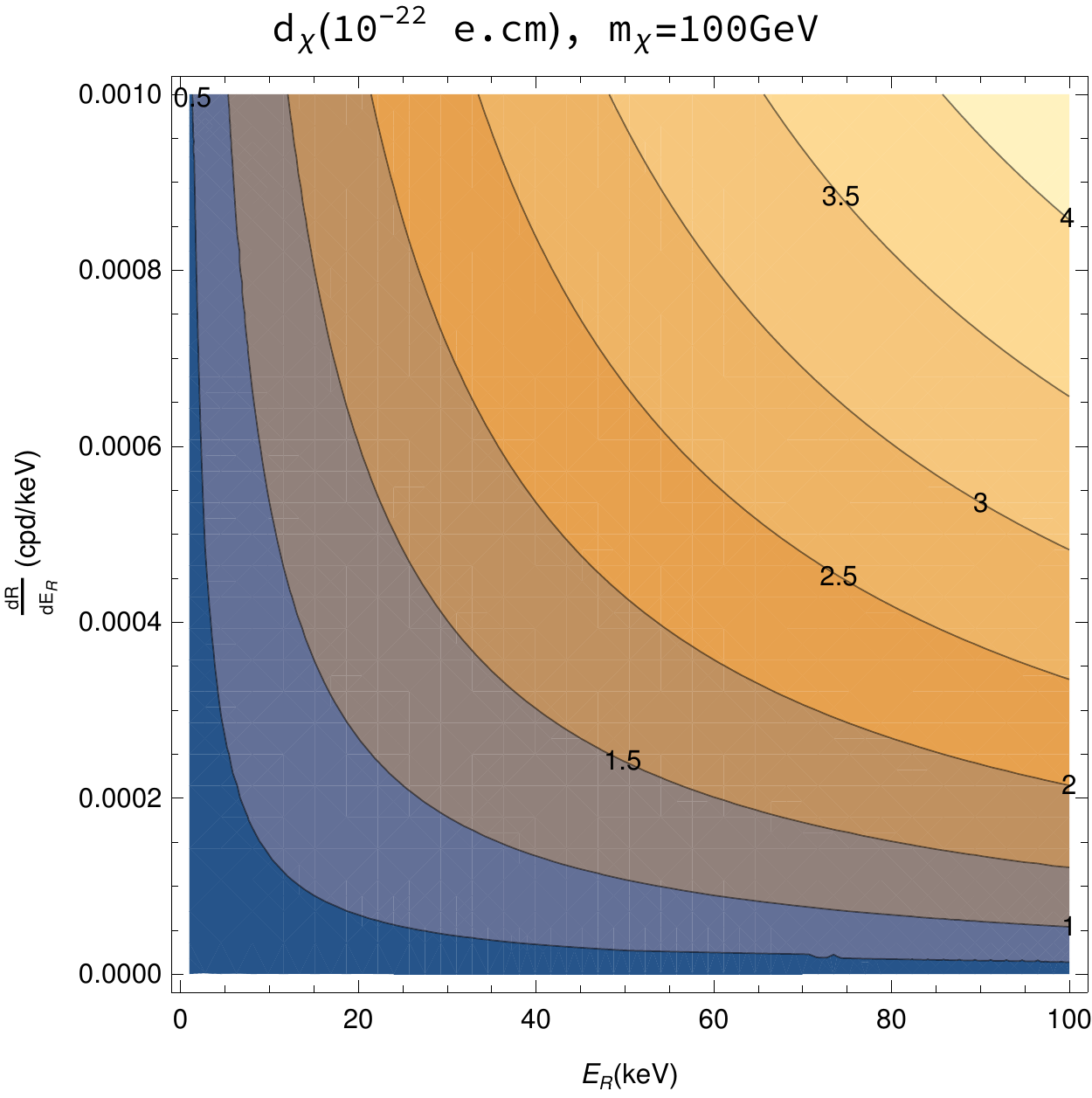}
		\caption{}
		\label{Fig4b}
	\end{subfigure}
	\caption{Electric dipole moment as as function of events rate/keV and the recoil energy for two values of $m_{\chi}=10$GeV (a) and $m_{\chi}=100$GeV (b)   }
	\label{Fig4}
\end{figure}

Similarly, in finding the bounds of electric dipole moment we set $\mu_{\chi}=0$ we can find bounds for $d_{\chi }$. 
In Figure \ref{Fig4} we illustrate the value of electric dipole moment as contour plot of the mass and recoil energy value. In case if the events/keV of order $\mathcal{O}(10^{-5})$ then for a mass $m_{\chi}=100$GeV the value of the electric dipole moment $d_\chi \geq 5 \times 10^{-23}e.cm$.\\

For a set of parameters we can find the maximal and minimal value for dipole moment. The minimal value is when we take into account the $Z, X$ bosons exchanges.
The maximal possible value of magnetic and electric dipole for this hidden Dirac particle is when there is no $Z, X$ bosons exchanges. Using the same set of parameters as above we obtain  $\mu_\chi^{max}=4.36 \times 10^{-9} \mu_B$ and $d_\chi^{max}=7.48 \time 10^{-24} e.cm$ for $\frac{dR}{dE_R}=10^{-5}$ and $m_\chi=10$GeV. In case $m_\chi=100$GeV we have   $\mu_\chi^{max}=1.37 \times 10^{-8} \mu_B$ and $d_\chi^{max}=2.36 \time 10^{-23} e.cm$. We can summarize our limits in the Table \ref{Table1}:

\begin{table}[ht]
	\centering
	
	{\begin{tabular}{ccc}
			\hline
			\multicolumn{3}{c}{$ \epsilon=0.003  ,\hspace{0.8mm} \hspace{0.8mm} m_X=10GeV, \hspace{0.8mm} g_X=0.1 ,\hspace{0.8mm} E_R=30keV$}\\
			[0.5ex] 
			\hline		
			 $m_\chi$(GeV) & $| \mu_{\chi}(10^{-8} \mu_B) |$ & $| d_{\chi}(10^{-23}e.cm) |$\\
			[0.5ex] 
			\hline
			10 & $  0.382101  \leq \mu_\chi \leq 0.435944 $ & $ 0.65586 \leq d_\chi \leq 0.74828$\\
			$100$ & $ 1.3626 \leq \mu_\chi \leq 1.37857$ & $ 2.33867 \leq d_\chi \leq 2.36626$\\	
			\hline
		\end{tabular}
		
		\caption{Bounds on the $\mu_{\chi}$ magnetic moment and $d_{\chi}$ electric  dipole moment}
		\label{Table1}}
\end{table}

Next, we will evaluate EDM and MDM based on the first results from the XENON1T \cite{XENON1T-Reach,XENON1T-Result,XENON1T-Analysis} dark matter experiment. The XENON1T experiment gives the events rate 1.5$\times 10^{-5}$ events/kg/day with recoil energy   $E_R \in [4.9,40.9 ] $ keV. The XENON1T gives limit for WIMP particle mass above 6GeV. Using the condition of positive value of magnetic dipole moment combine with XENONT1T limit, solving for zero value of magnetic dipole moment as function of DM mass $m_\chi$,  we can find possible range for $\epsilon$ and  for the mass of X boson. If $   0.003 \leq \epsilon \leq 0.01$ then $  5 \leq m_X \leq 9$GeV. This mass range of the gauge boson associated with the hidden $U(1)_X$ is in agreement with previous works \cite{Higss-Decay-Window}. In this paper, the exotic decay  of Higgs boson ($H \rightarrow X Z_d \rightarrow 4l$) is examined at LHC. In this decay channel,   $X=Z, Z_d, \gamma $ and $Z_d$ is the gauge boson of hidden $U(1)_d$. The mass of the hidden boson $Z_d$ is constrained to be in $ [5GeV, 10GeV]$. This mass range is within LHC detection ability for hidden sector \cite{LHC-probes-HS,ILC-HS}. The range of the kinetic mixing parameter $\epsilon$ is also possible in ability of ILC \cite{ILC-HS}.

Similar to previous section, the XENON1T also gives bounds of dipole moment for mass $m_\chi=100$ GeV are:   $ 1.93448 \times 10^{-8}\mu_B \leq \mu_\chi \leq 1.9496 \times 10^{-8}\mu_B$ and   $ 3.3204 \times 10^{-23}e.cm \leq d_\chi \leq 3.3464 \times 10^{-23}e.cm$.

\section{Conclusion}

The existence of a hidden $U(1)_X$ gauge symmetry has been studied recently in literature. By introducing new Dirac Dark Matter, this hidden Dirac DM can interact with the SM sector through kinetic mixing between $U(1)_X$ and $U(1)_Y$. The appearance of the hidden Dirac fermion also leads to the possible existence of the dipole moment of the hidden Dirac fermion.   In this work, we investigate the interaction of dark matter with nuclei through $\gamma $, $Z$ boson and the new $X$ boson from dark sector. We evaluate the event rate of hidden DM scatter off XENON nuclei and  set bounds for the mass of the hidden Dirac fermion and the value of the Dipole moment.  Using the current results for direct detection of dark matter at XENON1T, the bounds for hidden Dirac particle are found to be $ 1.93448 \times 10^{-8}\mu_B \leq \mu_\chi \leq 1.9496 \times 10^{-8}\mu_B$ and   $ 3.3204 \times 10^{-23}e.cm \leq d_\chi \leq 3.3464 \times 10^{-23}e.cm$ for mass $m_\chi=100$ GeV. From the condition of the existence of dipole moment of hidden dark matter,  we can constraint the kinetic mixing parameters $ 0.003 \leq \epsilon \leq 0.01$ and the mass of new hidden boson      5GeV$ \leq m_X \leq$ 9GeV.

\section*{Acknowledgment}
Acknowledgment
H. N. L and V. H. B. acknowledge the financial support of the Vietnam Academy of
Science and Technology under grant NVCC 05.03/20-20.\\[0.3cm]

\end{document}